\documentclass[preprint]{aastex}

\begin{document}

\title{The Period Variation of and a Spot Model for the Eclipsing Binary AR Bootis}
\author{Jae Woo Lee$^{1}$, Jae-Hyuck Youn$^{1}$, Chung-Uk Lee$^1$, Seung-Lee Kim$^1$, and Robert H. Koch$^2$}
\affil{$^1$Korea Astronomy and Space Science Institute, Daejeon 305-348, Korea}
\email{jwlee@kasi.re.kr, jhyoon$@$kasi.re.kr, leecu@kasi.re.kr, slkim@kasi.re.kr}
\affil{$^2$Department of Physics and Astronomy, University of Pennsylvania, Philadelphia, USA}
\email{rhkoch@earthlink.net}

\begin{abstract}
New CCD photometric observations of the eclipsing system AR Boo were obtained from February 2006 to 
April 2008. The star's photometric properties are derived from detailed studies of the period variability 
and of all available light curves. We find that over about 56 years the orbital period of the system has varied 
due to a combination of an upward parabola and a sinusoid rather than in a monotonic fashion. Mass transfer 
from the less massive primary to the more massive secondary component is likely responsible for at least 
a significant part of the secular period change. The cyclical variation with a period of 7.57 yrs and 
a semi-amplitude of 0.0015 d can be produced either by a light-travel-time effect due to an unseen companion 
with a scaled mass of $M_3 \sin i_3$=0.081 $M_\odot$ or by a magnetic period modulation in the secondary star. 
Historical light curves of AR Boo, as well as our own, display season-to-season light variability, 
which are best modeled by including both a cool spot and a hot one on the secondary star. 
We think that the spots express magnetic dynamo-related activity and offer limited support for preferring 
the magnetic interpretation of the 7.57-year cycle over the third-body understanding. Our solutions confirm 
that AR Boo belongs to the W-subtype contact binary class, consisting of a hotter, less massive primary star 
with a spectral type of G9 and a companion of spectral type K1. 
\end{abstract}

\keywords{binaries: close --- binaries: eclipsing --- stars: individual (AR Bootis) --- stars: spots}{}

\section{INTRODUCTION}

In the vicinity of the globular cluster M3, AR Boo (GSC 1999-0011; $V$=$+$12.76, $B$--$V$=$+$0.89) was announced 
by Kurochkin (1960) to be an eclipsing binary with an asserted period of 0.416718 d. Houck \& Pollock (1986) 
obtained a photographic light curve and determined a new period of 0.3447 d. Kurochkin's period turned out to 
be incorrect probably because there were only 97 magnitude estimates over a ten-year interval. Since then, 
CCD light curves have been made by Wolf et al. (1998) and by Samec et al. (2006) in the $VR$ and $BVRI$ bandpasses, 
respectively. The former authors reported that the orbital period history of the system can be represented either 
by an abrupt period increase about 1983 or by a positive quadratic term indicating a continuous period increase. 
Samec et al. analyzed their light curves using the Wilson-Devinney synthesis code (Wilson \& Devinney 1971, 
hereafter WD) and classified the binary as a member of the W-subtype cateory defined empirically by Binnendijk (1970). 
Such binaries show light curves in which the deeper minimum corresponds to an occultation of the smaller, less 
massive component by the larger, more massive one.  

In addition, these latter authors suggested that a second-order light variation is due to a single hot spot 
on the more massive component and that the orbital period is increasing. In this article, we present 
improved photometric properties of AR Boo from detailed studies of the period and light variations, based on 
our new CCD observations as well as on historical data.

\section{OBSERVATIONS}

We performed time-series CCD photometry of AR Boo on 27 nights between February 2006 and April 2008, 
using an FLI IMG4301E CCD camera and a $BVR$ filter set attached to the 1.0-m reflector at 
the Mt. Lemmon Optical Astronomy Observatory (LOAO) in Arizona, USA. The detecting system had been installed 
by the Korea Astronomy and Space Science Institute (KASI) and was operated by remote control from Daejeon, 
Korea via an internet connection (Han et al. 2005). The CCD chip has 2084$\times$2084 pixels and a pixel size 
of 24 $\mu$m. The image field-of-view is 22$\arcmin$.2$\times$22\arcmin.2 at the f/7.5 Cassegrain focus 
of the telescope. With the conventional IRAF package, we processed the CCD frames to correct for bias level, 
dark noise, and  pixel-to-pixel inhomogeneities of quantum efficiency and applied simple aperture photometry 
to obtain instrumental magnitudes for about a hundred stars near to the variable.

The nearby stars GSC 1999-0009 and GSC 1999-0067 (on the same CCD frame as the variable) were selected as 
comparison (C) and check (K) stars, respectively. For these stars no light variability had been reported 
in the previous observations by Samec et al. and we also detected no variations in the brightness difference 
between them during our observing runs. The observations of our first season were dedicated to obtaining 
new times of minimum light. From the 2007 and 2008 seasons, a total of 1838 individual observations was 
obtained in the three bandpasses (617 in $B$, 619 in $V$, and 602 in $R$) and a sample of them is listed 
in Table 1. The light curves are displayed in the upper panel of Figure 1 as magnitude differences {\it versus} 
orbital phase, where the open circle and plus symbols show the individual observations of the 2007 and 2008 seasons, 
respectively. The differences ('07-'08) between the two seasons are plotted in the middle panel. To obtain 
these seasonal differences, we interpolated in phase and magnitude difference between phase-consecutive measures 
of the 2008 season to the phase of an appropriate 2007 observation and performed the subtraction.  
All (K--C) magnitude differences appear in the lower panel. Whereas the reference stars were evidently constant, 
AR Boo has clearly varied seasonally.

\section{ORBITAL PERIOD STUDY}

From our CCD observations, times of minimum light in each filter were determined by the method of 
Kwee \& van Woerden (1956). Twenty-two weighted mean timings from these determinations and 84 epochs 
(12 photographic plate, 13 visual, 3 photographic, and 56 CCD) from the literature were used for a period analysis 
of AR Boo. Most of the eclipse timings were extracted from the data base referenced by Kreiner et al. (2001). 
All available CCD timings are listed in Table 2, where the second column gives the standard deviation of each timing. 
Because the other timings were published with no errors, the following standard deviations were assigned 
to timing residuals based on observational method: $\pm$0.0130 d for photographic plate, $\pm$0.0096 d for visual, 
and $\pm$0.0012 d for photographic minima. For ephemeris computations, relative weights were then calculated 
as the inverse squares of these values consistent with errors and weights for the CCD timings.

As previously mentioned, the period variation of AR Boo has been interpreted to be due to either a sudden period jump 
or a continuous period increase.  After testing several ephemeris forms, we found that the history of the period 
shows a short-term oscillation superposed on a upward parabolic variation, rather than varying in monotonic or 
bi-linear fashions. This type of period variation appears in contact and near-contact binaries such as 
ER Ori (Kim et al. 2003), BX Peg (Lee et al. 2004), TU Boo (Lee et al. 2007), and V432 Per (Lee et al. 2008), 
for which orbital period changes of the systems can be satisfactorily explained as combinations of secular variations 
and light-travel time (hereafter LTT) effects.  In general, there also exists the possibility that a binary orbit 
has shrunk from an initially wider dimension by secularly transferring the angular momentum of an inner, close pair 
to a distant companion of a triple system (cf. Pribulla \& Rucinski 2006). A recent study by these authors shows 
that about 42\% of 151 known contact binaries brighter than $V_{\rm max}=$ +10 exist in multiple systems.

We, therefore, initially assumed the oscillation to be due to an LTT effect and all eclipse timings were fitted 
to a quadratic {\it plus} LTT ephemeris as follows:
\begin{eqnarray}
C = T_0 + PE + AE^2 + \tau_{3},
\end{eqnarray}
where $\tau_{3}$ is the LTT due to a hypothetical distant companion to the eclipsing close binary (Irwin 1952, 1959)
and includes five parameters ($a_{12}\sin i_3$, $e$, $\omega$, $n$ and $T$). Here, $a_{12}\sin i_3$, $e$ and $\omega$ 
are the orbital parameters of the eclipsing pair around the mass center of the triple system. The parameters $n$ and
$T$ denote Keplerian mean motion of the mass center of the eclipsing pair and the epoch of its periastron passage,
respectively. The Levenberg-Marquart algorithm (Press et al. 1992) was used to evaluate the eight parameters of 
the ephemeris and the final results are summarized in Table 3, together with other derived quantities.  
The $O$--$C$ residuals calculated from the linear terms in equation (1) are plotted in the upper panel of Figure 2. 
The middle and lower panels display the CCD residuals from the linear {\it plus} quadratic term and from 
the complete ephemeris, respectively. The $O$--$C_{\rm full}$ residuals are given in the fourth column of Table 2. 
As displayed in Figure 2, the quadratic {\it plus} LTT ephemeris currently provides a good representation of 
all the $O$--$C$ residuals. This cyclical variation can be interpreted as an LTT effect due to an additional object 
with a mass of at least $M_3 \sin i_3$=0.081 $M_\odot$. The minimum mass of the third companion is close to 
the theoretical limit of about 0.072 $M_\odot$ for a brown dwarf star.  Because of noise in even the recent timings 
and also because only about 1.5 cycles of the 7.57-year period have been covered, these results are certainly preliminary.

This interpretation is not unique.  It is alternatively possible that a cyclical period modulation may be caused 
by a magnetic activity cycle in a binary system with at least one star having a spectral type later than about F5. 
Applegate (1992) and Lanza et al. (1998) have both discussed this possibility. The latter reference constrains 
and elaborates the concepts of the former one but both developments remains too conservative for a binary such 
as AR Boo.  They are limited to considering detached objects such as RS CVn and AR Lac and, for the present system, 
quadrupole gravitational moments do not express the complex mass distributions of the AR Boo stars.

If this concern is ignored, curve-fitting of the period residuals remains the same as for the mechanical model 
just described above, but no orbital meaning is attached to the shape and phasing of the curve.  In order to calculate 
model parameters for magnetic activity, we applied the period ($P_3$) and amplitude ($K$) for the cyclical variation 
to Applegate's formulae. The results are presented in Table 4, where $\Delta P$, $\Delta Q$ , $\Delta J$, $I_s$, 
$\Delta \Omega$, $\Delta E$, $\Delta L_{\rm rms}$, and $B$ denote the amplitude of orbital period modulation, 
the variation of the gravitational quadrupole moment, the rate of angular momentum transfer, the moment of inertia of 
the convective shell, the variable fraction of differential rotation, the work needed to transfer the angular momentum, 
the rms luminosity variation and mean subsurface magnetic field intensity, respectively. In addition to these paramaters,
the quantity $\Delta m_{\rm rms}$ denotes a bolometric magnitude difference relative to the mean light level of AR Boo 
converted to magnitude scale with equation (4) in the paper of Kim et al. (1997). The light variation ($\Delta L_{\rm rms1}$) 
ascribed to the hotter, primary star is a factor of 5 greater than the value ($\Delta L/L_{\rm {1,2}} \sim 0.1$) 
proposed by Applegate, while that ($\Delta L_{\rm rms2}$) for the secondary is close to his theoretical value. 
(Actually, Lanza et al. suggest that Applegate's limiting values may be relaxed considerably.)  These results indicate 
that a magnetic dynamo seated in the secondary star can also adequately explain the observed period modulations of 
AR Boo. Distinguishing between the mechanical and magnetic interpretations of the 7.57-year oscillation requires 
additional information, which will be suggested later after the light curve fittings.

In equation (1), the coefficient of the quadratic term is positive and represents a continuous period increase 
with a rate of $dP$/$dt$ = 2.68$\times$10$^{-7}$ d yr$^{-1}$. It is conventional to ascribe such a variation to
conservative mass transfer between the stellar components in the system. The transfer rate from the less massive star
to the more massive component of AR Boo is calculated to be about 1.48$\times$10$^{-7}$ M$_\odot$ yr$^{-1}$ 
from the relation between $dP$/$dt$ and the masses of the component stars calculated later. This value is larger than  
those recently derived for other W-subtype contact systems [e.g., 1.3$\times$10$^{-7}$ for AD Cnc (Qian et al. 2007), 
2.9$\times$10$^{-8}$ for TY Boo (Yang et al. 2007), and 2.3$\times$10$^{-8}$ for CC Com and 2.0$\times$10$^{-8}$ 
for BV Dra (Yang et al. 2009), all in M$_\odot$ yr$^{-1}$ units] and is a generally accepted one 
for classical semi-detached Algols (cf. Richard \& Albright 1999).  One might imagine that the calculated $dP$/$dt$
could be apportioned between (a) mass transfer from the hot to the cool eclipsing member and (b) angular momentum transfer 
to the possible third companion but there is no independent support for this idea.

\section{LIGHT VARIATIONS AND SPOT MODELS}

As do previous light curves of AR Boo, the new ones show continuously curving eclipses and also 
conspicuous seasonal light variations.  Within about 0.004 mag, the light curves of 2007 present equal light 
levels at the quadratures (Max I and Max II) while those of 2008 show the inverse O'Connell effect with Max I 
fainter than Max II by about 0.028, 0.023, and 0.018 mag for the $B$, $V$, and $R$ bandpasses, 
respectively. Moreover, the light maxima are slightly displaced to orbital phases 0.264 and 0.745 in 2007 and 
0.257 and 0.745 in 2008. These features are usually indicative of spot activity on the component stars 
and the striking seasonal light changes in AR Boo indicate that that activity has varied significantly with time.

Photometric solutions of AR Boo have been reported only by Samec et al. (2006) and no spectroscopic observations 
have been published for the system. We analyzed our data in a manner similar to that for the cool binaries TU Boo 
(Lee et al. 2007) and V432 Per (Lee et al. 2008). Our light curves were normalized to unit light at phase 0.75 
and were solved by using contact mode 3 of the WD code. The surface temperature of the more massive star was 
fixed at 5100 K from Flower's (1996) table, according to ($B-V$)=$+$0.886$\pm$0.010 given by Samec et al. 
and $E$($B-V$)=$+$0.015 calculated following Schlegel et al. (1998). The logarithmic bolometric ($X$, $Y$) 
and monochromatic ($x$, $y$) limb-darkening coefficients were interpolated from the values of van Hamme (1993) 
and were used in concert with the model atmosphere option. To avoid possible confusion, we refer to the primary 
and secondary stars as those being eclipsed at Min I and Min II, respectively. 

We first conducted a detailed $q$-search procedure over the range 0.3 $\le q \le$ 3.8. In this process,
we did not consider either a third light source ($l_3$) or spots. The search was made simultaneously 
for all light curves of each season and the results are plotted in Figure 3. From the figure, we can see that 
the $q$-search result of 2007 is different from that of 2008 and that local minima occur around $q$=1.3 and 
$q$=2.7 in 2008. This may be caused by the year-to-year variability of the system and the noisiness of 
the 2008 light curve. Because of this confusion, the $q$-search of the 2008 observations is considered 
indeterminate. We did find that the light-curve synthesis in 2007 is slightly better if $q>$1.0 and 
that an acceptable solution is close to $q$=2.6. This indicates that AR Boo is an W-subtype contact binary. 
To obtain an unspotted solution, we analyzed all six light curves simultaneously by treating the initial value 
of $q$ as an adjustable parameter. The results are listed in Table 5 and the residuals from the fitting are 
plotted in the left panels of Figure 4. As indicated by these panels, the computed light curves 
do not fit our data satisfactorily.  

In many contact binaries, second-order light variations have been attributed to local atmospheric inhomogeneities 
that may result from either magnetic dynamo-related activity or mass transfer between the component stars. 
Because the second-order light effects are themselves seasonally variable for AR Boo, they cannot be caused 
by a gas stream from the less massive primary impacting the inner hemisphere of the more massive secondary component 
unless the flow rate were itself variable. Also, our period study suggests checking for a third light source 
in the system. Accordingly, we re-analyzed our light curves by considering both a spot and such a source. The result 
is given as the one-spot model of Table 6 and the light residuals from the model are plotted in the middle panels 
of Figure 4. From these displays, we see that a single cool spot on the larger secondary component with 
a third light source does improve the light-curve fitting greatly but that large discrepancies still remain in 
the first half-period of the 2007 data. This detail can reasonably be modeled by an additional hot spot on 
the surface of the secondary.  It remains an open question whether such a spot results from mass transfer or 
is an intrinsic feature of the cool star. The final result is listed as the two-spot model of Table 6 and 
the residuals from this binary model are plotted in the right panels of Figure 4. Our solutions for AR Boo show 
that $l_3$ contributes 2.0-2.8 \% light in $B$, 1.2-1.5 \% in $V$, and 0.9-1.0 \% in $R$.  These values seem to 
point to a source relatively hotter than the binary components but the problem with this interpretation is that 
the $l_3$ values are not individually statistically significant.  A final light curve fitting with $l_3$=0 
led to parameter evaluations which, within the errors, did not change from those given in the table.  
The second-order light variations are best fitted by the simultaneous existence of a cool spot and a hot spot 
on the cool secondary star.  It should be noticed, however, that the two-spot model of 2008 gives a value of 
$\Sigma W(O-C)^2$ identical to that for the one-spot model and does not improve the light-curve fitting appreciably.  
Although the effect is not clearly seen in Fig. 1, the 2008 light curves are appreciably noisier than those of 
the preceding year and the difference is evident in all the panels of Fig. 4.  This difference makes 
a large contribution to the unchanged fitting criterion for 2008 and also to the large errors associated with 
the 2008 hot spot. Although we could have ignored that spot in 2008, we decided to leave its formal determination 
just as the WD code found it.

To take the understanding of AR Boo further and to examine whether our solutions can reasonably describe
the Samec et al. light curves, we analyzed them using our photometric parameters as initial values. The results 
are given in columns (2)-(3) of Table 7 and plotted in Figure 5 as continuous curves. Within the errors 
yielded by the WD code, the stellar parameters from the Samec et al. data are in satisfactory accord with those 
from our light curves. We see again that the formal $l_3$ values are not significant and that the differences 
among the (now 3) seasonal light curves can be explained almost entirely by spot variability, especially by 
the longitude drifts of the spots with time. 

The light curves of Wolf et al. (1998) do not have the observational weight of ours and we held only 
limited expectations for their study.  Since the Samec et al. and our light curves had converged to 
consistent geometric and photometric parameters, we assumed these values to be appropriate for the Wolf et al. ones 
and analyzed these latter only for spot descriptions. The results are listed in columns (4)-(5) of Table 7 and 
lead to the continuous curves in Figure 6.  These fit the observations satisfactorily.  Nonetheless, 
as in modelling the 2008 light curves, the Wolf et al. data do not need a hot spot, which may be a main cause 
of the large errors for the spot parameters.

The 4-season mean values of $l_3$$\approx$ 2 \%, if real, could come from a third object either gravitationally 
bound to or only optically related with the eclipsing pair AR Boo. We do not see such an object optically resolved 
on our time-series CCD images. Further, AR Boo is at a very high galactic latitude so background or 
foreground contamination is not likely. A previous section of this report quantified a possible projected orbit for 
a bound third body. If that supposed object be identified with the ``third light" source, it would have to be 
more than 3 magnitudes fainter than the hot eclipsing component and somewhat bluer than that star, i.e. a sub-dwarf.  
The evidence of the errors in Tables 6 and 7 forces the simpler understanding: the ``third light" values are 
only artifacts of the light curve fitting process.  If such an object exists, it has not yet been photometrically 
identified from all these light curves.

\section{DIMENSIONED STELLAR PARAMETERS}

The more massive secondary component with a spectral type of K1 V was assumed to conform to Harmanec's (1988) 
relation for masses and radii as functions of spectral type.  The remaining absolute parameters were then derived from 
the two-spot model of 2007 in Table 6.  All the stellar characteristics are listed 
in Table 8 wherein the bolometric corrections were obtained from the relation between $\log T$ and BC given by 
Kang et al. (2007). The apparent visual magnitude $V$=+12.76 at Max II (Samec et al. 2006) and our computed light ratio 
lead to $V_1$=+13.87, and $V_2$=+13.27 for the primary and secondary stars, respectively. The distance to the system 
was calculated to be about 350 pc from the interstellar absorption of $A_V$=0.05 and the values of $V_2$ and $M_{V_2}$. 

The masses and radii in Table 8 are not close to those of main sequence stars but they do conform to the general
pattern of contact binaries shown long ago by Hilditch et al. (1988) when it is remembered that those authors
define primary and secondary stars by mass and not by temperature.  If the assumption is made that the hot star 
is a normal main sequence one, the sum of the masses is unrealistically large. 
Figure 3 shows that there can be considerable uncertainty in determining the mass ratio as was noted in that section
of the text.  With a presently unknowable impact, this uncertainty propagates into the geometrical parameters of the 
synthesis and ultimately into the stellar radii and masses.  Quite likely, radial velocities would diminish this 
uncertainty since the spectrum should be double-lined if the light ratio is correct.

\section{THE SPOTS AND THE 7.57-YEAR PERIOD VARIATION}

This paper gives a summary of the discontinuous history of light curve morphology of AR Boo from 1997 into 2008.  
One description of the second-order morphology emphasizes that cool spots have manifested themselves over 
a siderographic band from $+89^\circ$ to $+160^\circ$ in longitude and 80$^\circ$ to 92$^\circ$ in co-latitude 
on the cool star.  Simultaneously and in the same coordinate sense, hot spots have been seen over the band 
$-62^\circ$ to $+65^\circ$ and 99$^\circ$ to 115$^\circ$ on the same star. Whether or not these hot spots are 
chromospheric plages in the solar sense is not known for AR Boo but we will use that noun here even if they 
turn out to be photospheric features. With respect to the stellar equator, the cool spots and plages have appeared 
in opposite hemispheres but close to and nearly parallel to that great circle. When account is taken of errors, 
the co-latitude excursions have been so small that the features may be said to have shown no latitude drift at all. 
The longitudes of the disturbed regions, on the other hand, have shown significant changes with time. 

Solar-type magnetograms cannot be made for an unresolved star and a very large telescope would be needed to
measure the normalized circular polarization component for AR Boo.  One must seek surrogates for 
these conventional indicators of a magnetic environment and the light curves are the only source of 
other markers. 

It is a reasonable working inference that the spots and plages are surface manifestations of 
an embedded magnetic field and activity consistent with the second understanding of the 7.57-year period variation.  
Possible quantifications of this activity might be (a) the fluxes emitted by either or both of the hot and 
cool spots, (b) the sum of these fluxes and (c) the ratio of the fluxes.  From the entries in Tables 6, 7 and 8 
the appropriate parameters have been extracted so as to calculate the spot and plage fluxes and their sums and 
ratios on the assumption of black body emissivity. The Samec et al. and our light curves fall upon 
the decreasing-period branch of the 7.57-year waveform and the models of these light curves show ordered relations 
for 3 of the criteria with time and, more suggestively, with amplitude in that waveform as follows. With advancing time, 
the progressively more negative excursions in the waveform are associated with (a) increases in the plage fluxes 
and decreases in the cool spot fluxes, (b) no monotonic change in the flux sum and (c) decreasing values of 
(plage flux/cool spot flux). (In the formulation by Lanza et al. these points, falling as they do on 
the negative slope fraction of the 7.57-year cycle, signal an increase of the cool star's quadrupole moment with 
concomitant changes in the magnetic dynamo.) The weakness of the associations is evident:  the spot characterizations 
from Wolf et al. do not conform to the relation defined by only the 3 points sampling the first half of the following 
cycle but, it is possible to extenuate this seeming failure in the following way.  One does not know what would be 
the appearance of the entire cycle of spot and plage activity as a function of the 7.57-year oscillation. Would it be
a simple closed oval, a degenerate single curve, or a twisted closed curve?
Perhaps then, the failure is only a seeming one and is really evidence that magnetic activity changes 
from one 7.57-year cycle to the next in the same manner that solar magnetic activity is not a constant of nature.  
If this would be so, there would also be no sense to the propagation of that cycle forward or backward in time as 
we have done in the middle panel of Figure 2.

\section{THE MINIMUM TIMINGS IMPROVED}

It may be imagined that the cyclical component of the period variability, discussed first in Section 3, could be produced 
by apparent phase shifts of the real conjunctions due to asymmetrical eclipse minima originating from starspot activity 
and/or the method of measuring the timings of minima (cf, Lee et al. 2004, 2008). To examine this possibility, 
we calculated anew the minimum epoch for each eclipse curve of Wolf et al. (1998), Samec et al. (2006), and LOAO 
with the WD code by means of adjusting only the ephemeris epoch ($T_0$) and not allowing a phase shift as in Table 6. 
The results are listed in the second column of Table 9 together with the previously-tabulated timings for comparison 
and are illustrated with the plus symbols in Figure 2. We can see that the differences among them are  
smaller than the observed amplitude (about 0.003 d) of the 7.57-year variation which therefore survives after spot 
correction.

There are, moreover, systematic runs with time between the observed timings and the WD ones.  In our data of 
Table 9, the column 4 differences in 2007 are positive for both minima, while those in 2008 are positive 
for Min I except for one eclipse (HJD 2454555.8471) and negative for Min II. In 2007, a cool spot and a hot spot are 
displayed to the observer before and after the first quadrature, respectively, which cause the positive values 
for both eclipses.  In 2008, however, a cool spot is seen at the first quadrature. This spot produces positive values 
for Min I but negative values for Min II.

Both the diminution in scatter among the timings and our spot interpretation of the timing displacements can be 
considered consistent with the expectation of Maceroni \& van't Veer (1994) that minimum timings are systematically 
shifted by light curve asymmetries due to starspot activity. Our results lend weight to the proposition that 
the light curve synthesis method leads to better timing determinations than does the Kwee \& van Woerden method 
that has been almost universally applied to individual eclispses for more than 50 years. 
This work on AR Boo presents the most convincing confirmation in the current literature of the predictions of 
those authors. Our interpretation also implicitly expresses the observational precision and the time density of 
measures that have to be attained so as to prevent spot effects from contaminating Keplerian ephemerides.  
Only if this be done can one seek out higher-order period variations. We suggest that all future period studies 
of this star use the timings of column 2 of Table 9 rather than those of Table 2. There is another recognition 
to be made. It will be impossible in general to correct casually-observed minimum timings for spot bias 
so some small systematic error will afflict the results of dedicated minimum monitoring programs.

\section{SUMMARY AND DISCUSSION}

In this article, we have analyzed new CCD observations of the eclipsing binary system AR Boo obtained during 
three successive seasons beginning in 2006 February.  The light curves show partial eclipses and 
season-to-season light variability. Our light-curve solution indicates that AR Boo is a W-subtype contact binary 
composed of a hotter G9 primary star and a cooler K1 companion, The derived spectral types depend, of course, on 
the adopted temperature for the secondary component and the observed ($B-V$) color index. 
The disturbed light curves of the system were best modeled by using a two-spot model with cool and hot spots 
on the secondary star. We speculate that both spots may be produced by magnetic dynamo-related activity.  

According to the thermal relaxation oscillation (TRO) theory (Lucy 1976; Lucy \& Wilson 1979), contact binaries 
must oscillate cyclically between contact and non-contact conditions. Because our detailed study of period and 
light variations show that the orbital period is increasing and that mass is transferring from the less massive primary 
to the more massive secondary star, AR Boo may presently be in a transition state evolving from a contact 
to a non-contact configuration as suggested by this theory. 

Our new orbital period study with all available timings reveals that the $O$--$C$ residuals have varied in 
a cyclical oscillation superposed on a secular period increase.  The semi-amplitude and period of 
the cyclical variation are low and short, respectively, and their values command modest confidence 
because the observed CCD timings cover only about 1.5 cycle of this period. The oscillation may be produced, 
in principle, either by the LTT effect due to a stellar- or substellar-mass companion of 
$M_3 \sin i_3$=0.081 $M_\odot$ or by a active magnetic cycle in the more massive secondary component
but not by the asymmetries of eclipse light curves due to starspot activity.  The circumstance that 
there has existed even a limited association between phase in the 7.57-year cycle and the spot activity 
seen in the light curves diminishes confidence in the postulated 3rd body while suggesting 
that a more searching attempt be made to understand magnetic activity mediating angular momentum exchange 
between the spin of the stellar mass distribution and the orbit.  

There's a further message to be drawn here.  Three of the authors had never even heard of this binary 
before the LOAO observations were in train. It is a very nondescript object, yet the assiduous accumulation 
of the best possible observations has given weight to two ideas:  starspots measurably inflect eclipse timings 
and magnetic cycles in stellar components of close binaries express themselves dynamically. Every binary has 
its individual information encoded for us but, a priori, we have no idea of the richness of that information and we must 
uncover it and piece it together with information from all the others that we observe.

In a perfect world there would be seasonal accumulation of light curves (and radial velocities) of AR Boo over 
the next 8 years and then surely understanding of this system and its activity would be advanced. It is even 
possible that the declination of the binary permits accumulation of two independent light curves per observing season.  
A justification for such a strenuous effort may be found in looking back at the 2007 and 2008 light curves.  
Whereas those of the earlier year were compiled over the 6 nights from March 11 through March 17, the data for 2008 
were observed over 17 nights from December 23, 2007 through April 12, 2008.

\acknowledgments{ }
We wish to thank Professor Chun-Hwey Kim for his help in using the $O$--$C$ database of eclipsing binaries. 
We also thank the staff of Mt. Lemmon Optical Astronomy Observatory for assistance with our observations. 
This research has made use of the Simbad database maintained at CDS, Strasbourg, France.

\clearpage
\begin{figure}
 \includegraphics[]{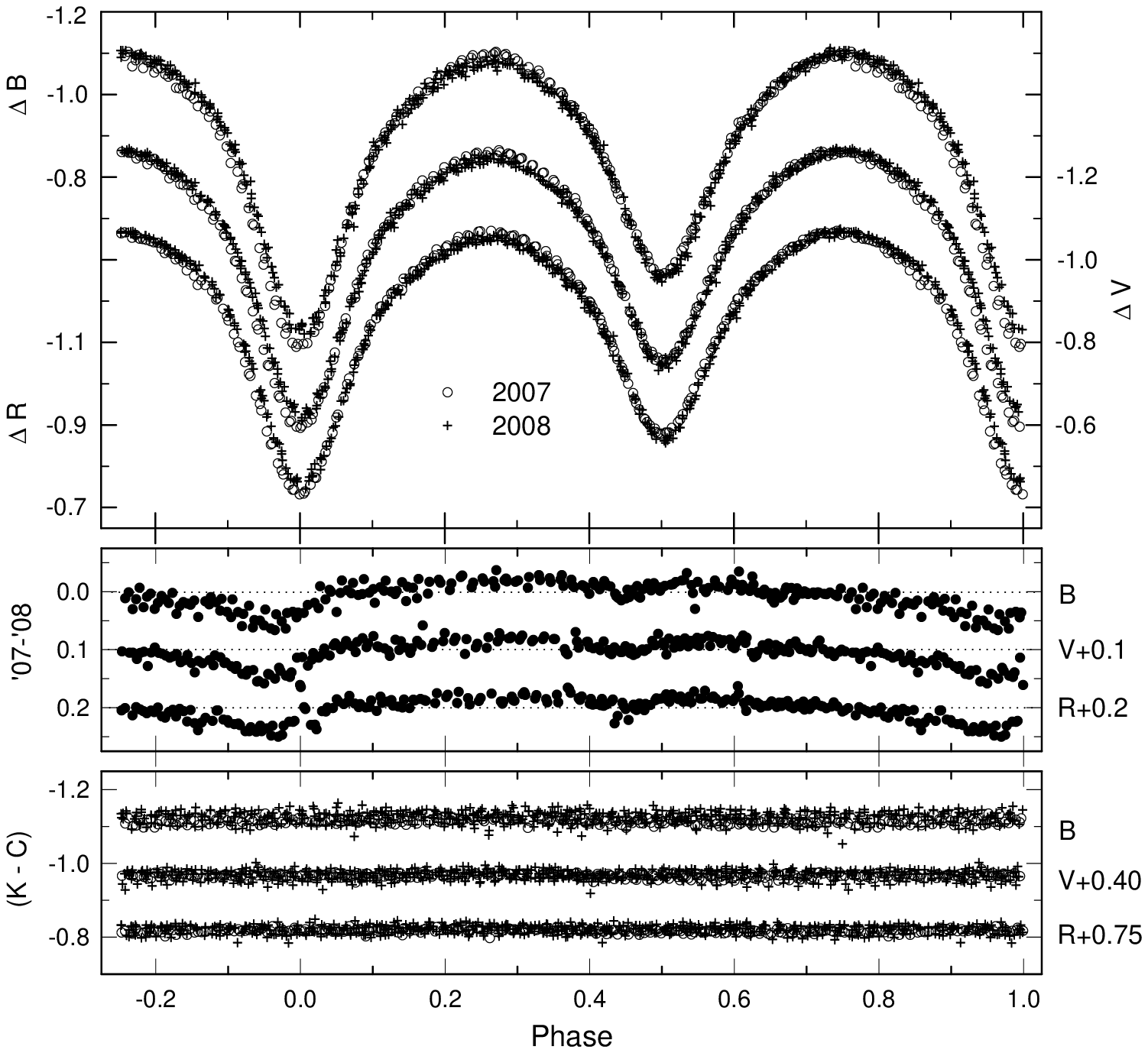}
 \caption{The upper panel displays our light curves of AR Boo in the $B$, $V$, and $R$ bandpasses.
 The differences between the two seasons are shown in the middle panel and the magnitude difference
 between the check and comparison stars in the lower panel. The open circle and plus symbols 
 in the upper and lower panels are the individual measures of the 2007 and 2008 seasons, respectively.
 The dotted lines in the middle panel refer to values of 0.0 mag.}
 \label{Fig1}
\end{figure}

\begin{figure}
 \includegraphics[]{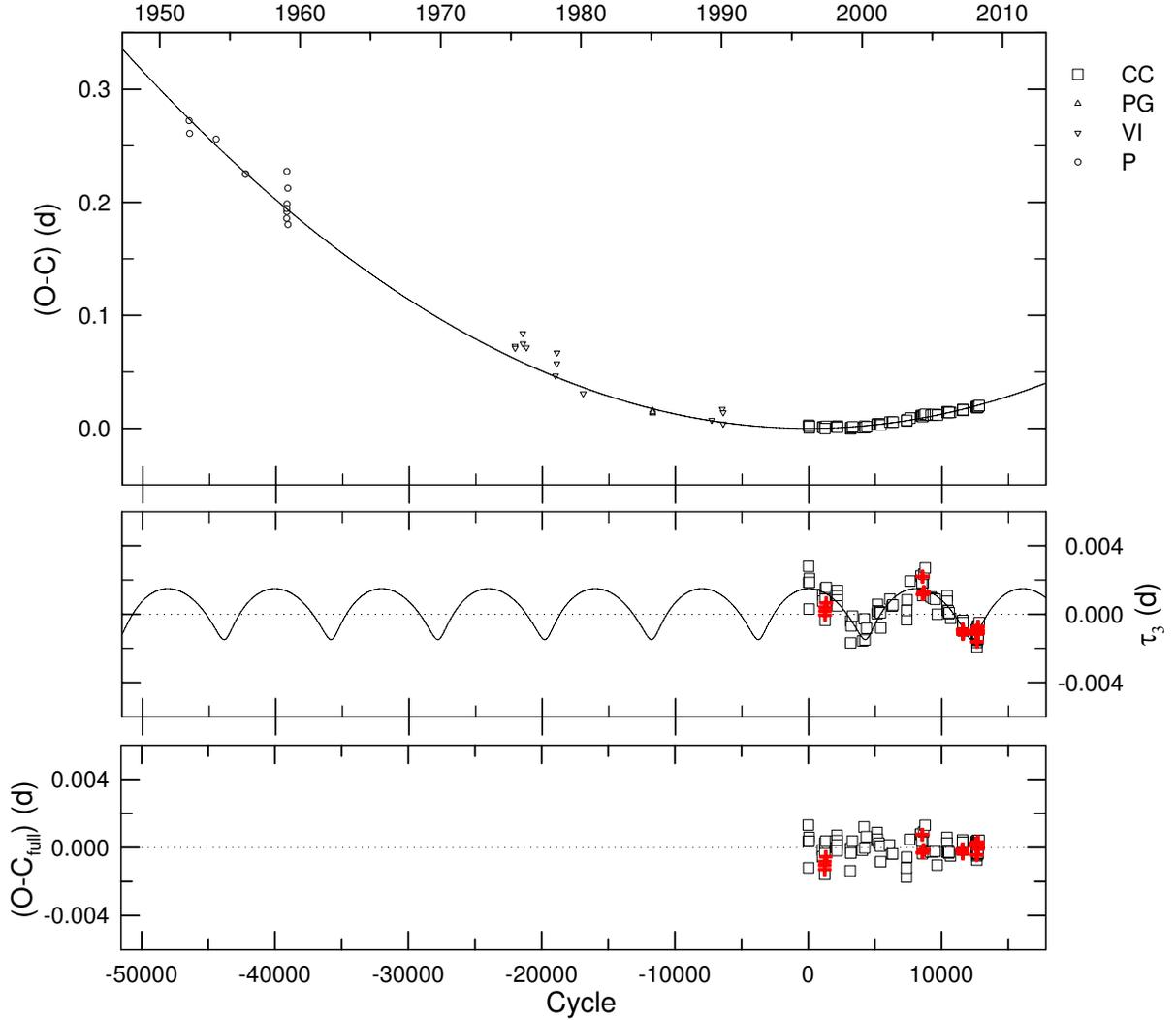}
 \caption{The $O$--$C$ diagram of AR Boo. In the upper panel the continuous curve represents 
 the final improved ephemeris. CC, PG, VI, and P stand for CCD, photographic, visual, and photographic plate minima, 
 respectively. The middle and lower panels display the CCD residuals from the linear and quadratic terms and 
 from the complete ephemeris, respectively. The 7.57-year cycle is shown propagated back in time only to indicate 
 that it could not have been recognized in the noisy earlier minimum timings. In the two lower panels, plus symbols 
 represent the minimum times obtained by re-analyzing individual eclipse curves with the WD code. } 
 \label{Fig2}
\end{figure}

\begin{figure}
 \includegraphics[]{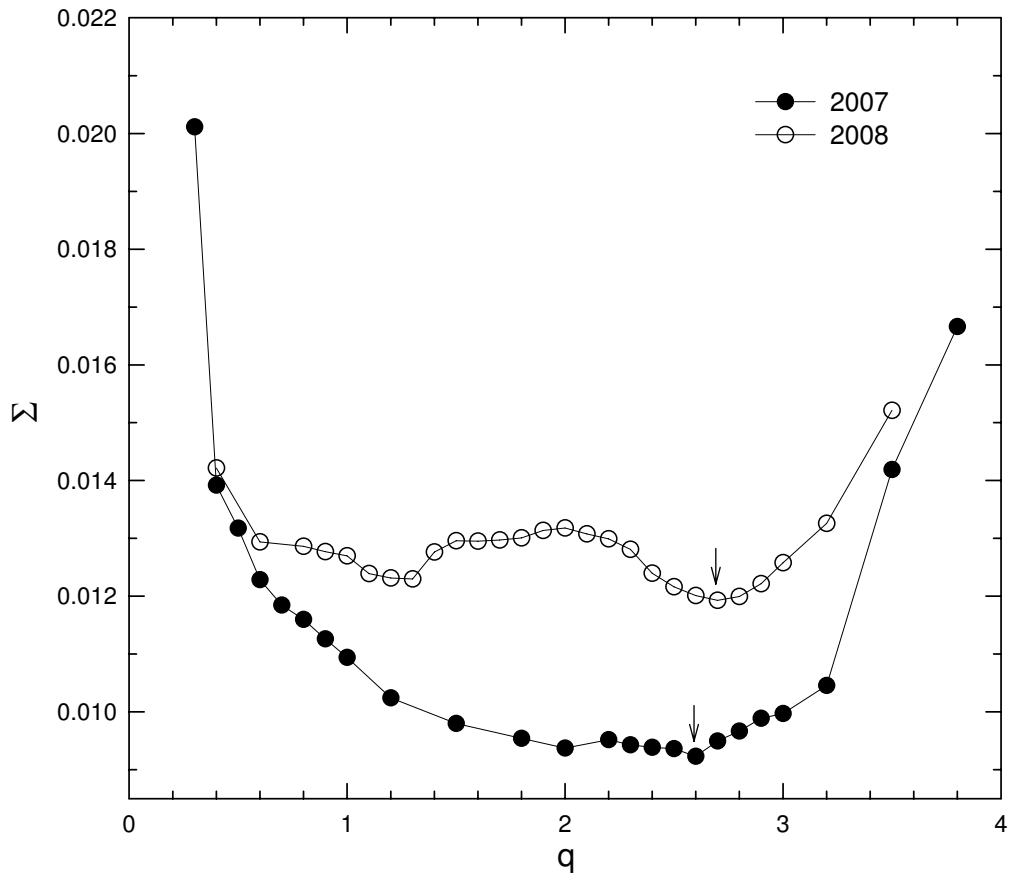}
 \caption{The behavior of $\Sigma$ (the sum of the residuals squared) of AR Boo as a function of mass ratio $q$. 
 The filled and open circles represent the $q$-search results for the 2007 and 2008 data sets, respectively. 
 The arrows indicate minimum values for each data set.}
\label{Fig3}
\end{figure}

\begin{figure}
 \includegraphics[]{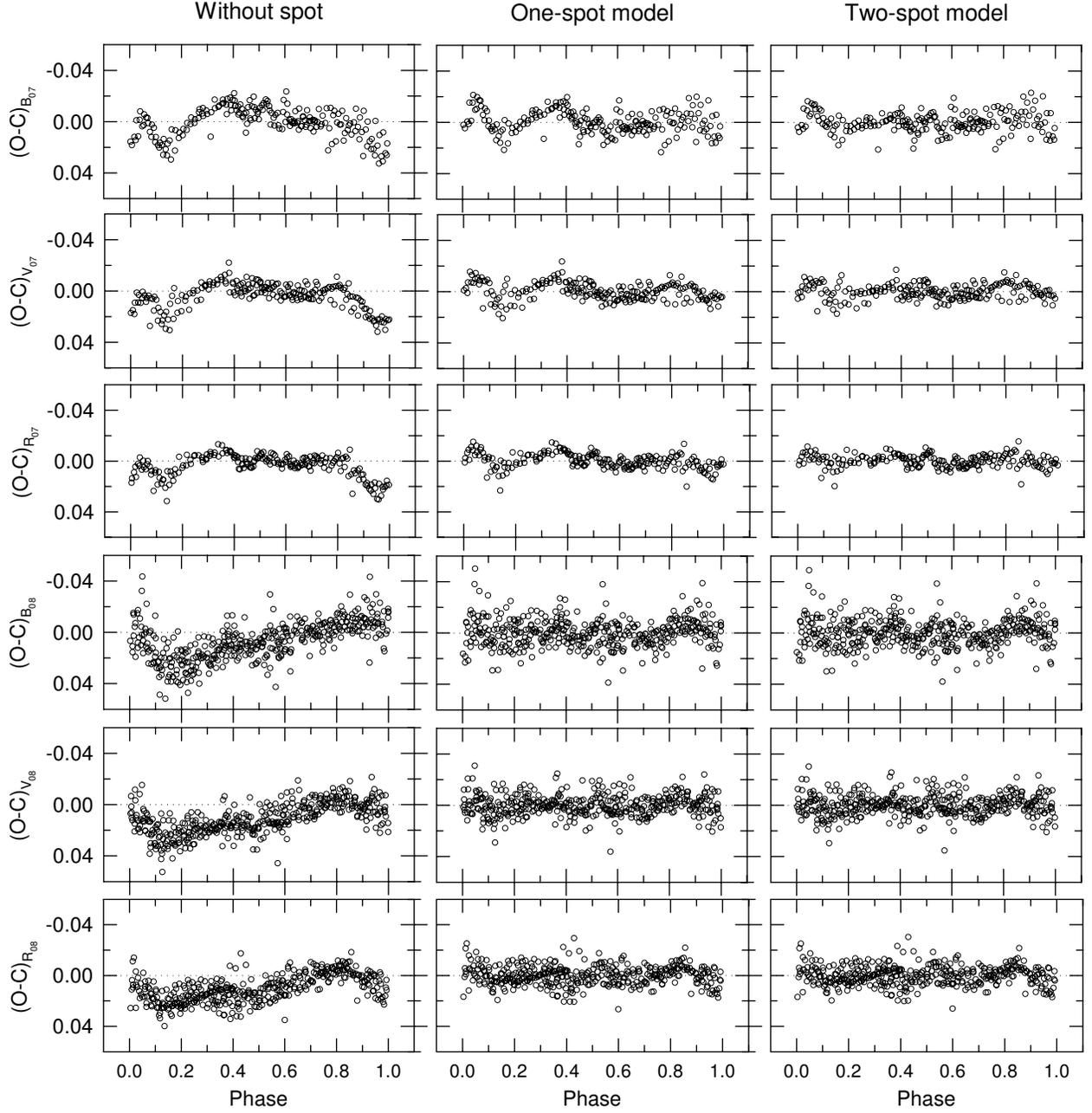}
 \caption{The left, middle, and right panels show the LOAO residuals from the solutions without a spot,
 a one-spot model, and a two-spot model listed in Tables 5 and 6, respectively.}
\label{Fig4}
\end{figure}

\begin{figure}
 \includegraphics[]{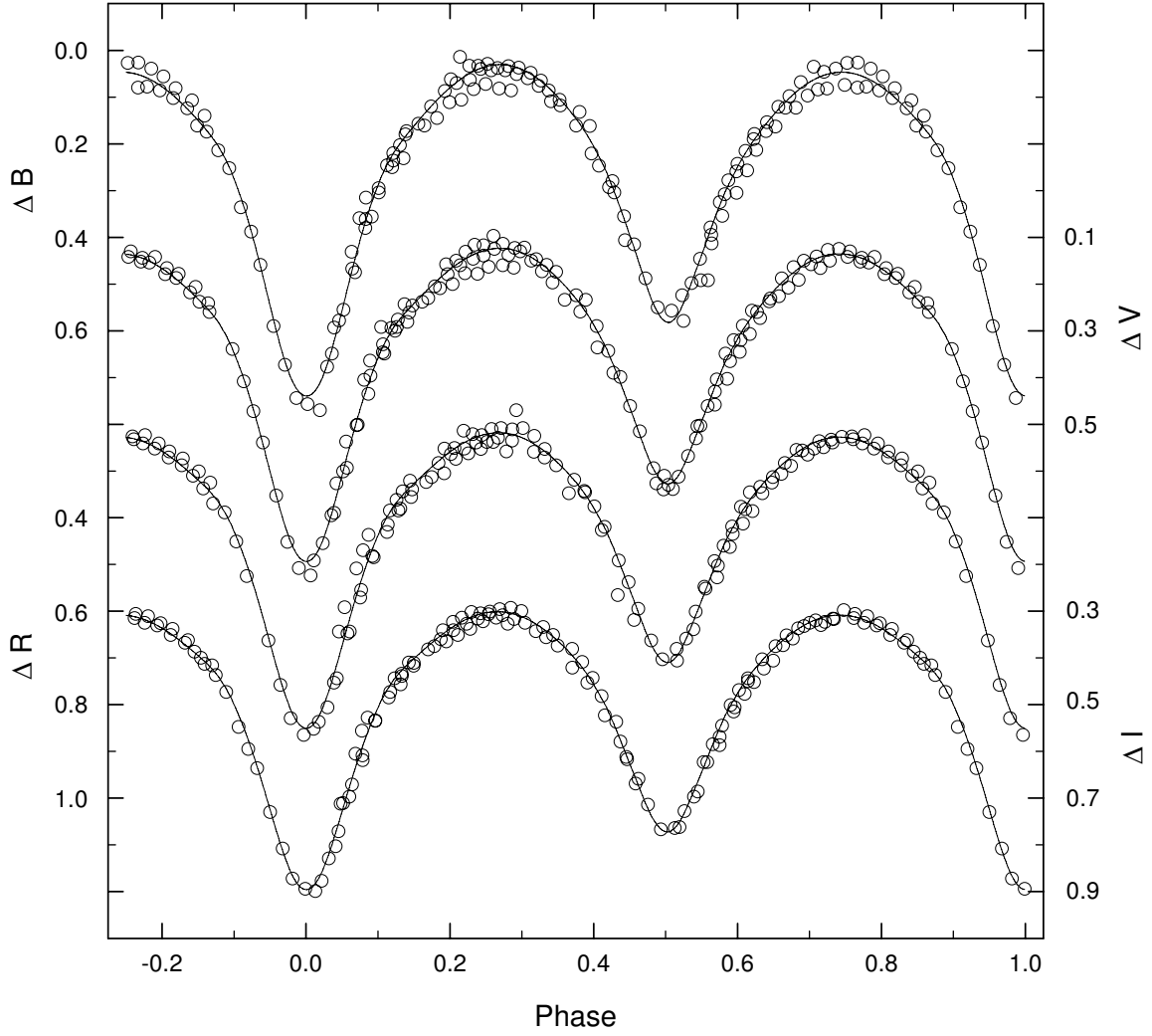}
 \caption{Light curves of Samec et al. (2006) in $BVRI$ bandpasses. The solid curves represent 
 the solutions obtained with the model parameters listed in columns (2)-(3) of Table 7.}
 \label{Fig5}
\end{figure}

\begin{figure}
 \includegraphics[]{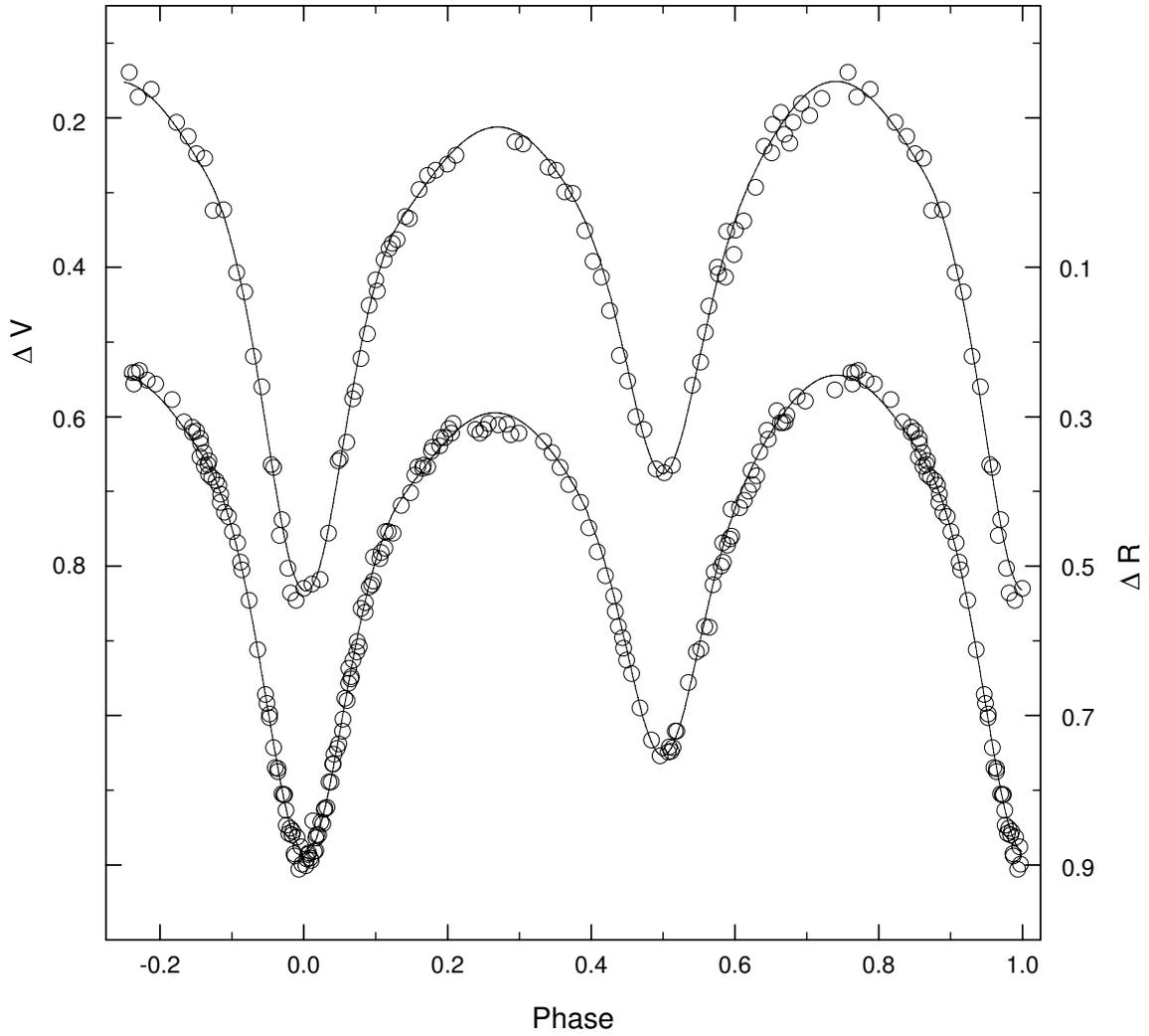}
 \caption{Light curves of Wolf et al. (1998) in $VR$ bandpasses. The solid curves represent 
 the solutions obtained with the model parameters listed in columns (4)-(5) of Table 7.}
 \label{Fig5}
\end{figure}

\clearpage
\begin{deluxetable}{cccccc}
\tablewidth{0pt} \tablecaption{CCD photometric observations of AR Boo.}
\tablehead{
\colhead{HJD} & \colhead{$\Delta B$} & \colhead{HJD} & \colhead{$\Delta V$} & \colhead{HJD} & \colhead{$\Delta R$} 
}
\startdata
2,454,171.79150 & $-$0.919  &  2,454,171.79340 & $-$1.063  &  2,454,171.79482 & $-$1.173   \\
2,454,171.79648 & $-$0.866  &  2,454,171.79842 & $-$1.026  &  2,454,171.79984 & $-$1.118   \\
2,454,171.80151 & $-$0.817  &  2,454,171.80350 & $-$0.972  &  2,454,171.80501 & $-$1.059   \\
2,454,171.80671 & $-$0.746  &  2,454,171.80872 & $-$0.912  &  2,454,171.81020 & $-$1.000   \\
2,454,171.81192 & $-$0.684  &  2,454,171.81392 & $-$0.841  &  2,454,171.81541 & $-$0.938   \\
2,454,171.81712 & $-$0.624  &  2,454,171.81913 & $-$0.782  &  2,454,171.82061 & $-$0.899   \\
2,454,171.82232 & $-$0.578  &  2,454,171.82432 & $-$0.763  &  2,454,171.82582 & $-$0.879   \\
2,454,171.82754 & $-$0.567  &  2,454,171.82956 & $-$0.758  &  2,454,171.83105 & $-$0.893   \\
2,454,171.83276 & $-$0.594  &  2,454,171.83476 & $-$0.792  &  2,454,171.83626 & $-$0.926   \\
2,454,171.83797 & $-$0.646  &  2,454,171.83998 & $-$0.846  &  2,454,171.84147 & $-$0.988   \\
\enddata
\tablecomments{This table is available in its entirety in machine-readable and Virtual Observatory (VO) forms 
in the online journal. A portion is shown here for guidance regarding its form and content.}
\end{deluxetable}

\begin{deluxetable}{lrrrcl}
\tabletypesize{\scriptsize}
\tablewidth{0pt}
\tablecaption{Observed CCD timings of minimum light for AR Boo.}
\tablehead{
\colhead{HJD} & \colhead{Error} & \colhead{Epoch} & \colhead{$O$--$C_{\rm full}$} & \colhead{Min} & References \\
\colhead{(2,450,000+)} & & & & & }
\startdata
0,180.4108  &  $\pm$0.0028  &     -6.0  &   0.00131  &  I   &  Saf\'ar \& Zejda (2000)    \\
0,182.4791  &  $\pm$0.0002  &      0.0  &   0.00036  &  I   &  Wolf et al. (1998)         \\
0,192.4789  &  $\pm$0.0030  &     29.0  &  -0.00119  &  I   &  Diethelm (1996)            \\
0,200.4128  &  $\pm$0.0002  &     52.0  &   0.00060  &  I   &  Wolf et al. (1998)         \\
0,200.5850  &  $\pm$0.0005  &     52.5  &   0.00037  &  II  &  Wolf et al. (1998)         \\
0,547.3555  &  $\pm$0.0002  &   1058.0  &  -0.00015  &  I   &  Wolf et al. (1998)         \\
0,551.4936  &  $\pm$0.0018  &   1070.0  &  -0.00054  &  I   &  Diethelm (1997)            \\
0,607.3635  &  $\pm$0.0004  &   1232.0  &  -0.00025  &  I   &  Wolf et al. (1998)         \\
0,607.5346  &  $\pm$0.0003  &   1232.5  &  -0.00159  &  II  &  Wolf et al. (1998)         \\
0,611.5025  &  $\pm$0.0005  &   1244.0  &   0.00026  &  I   &  Wolf et al. (1998)         \\
0,638.4028  &  $\pm$0.0002  &   1322.0  &   0.00038  &  I   &  Wolf et al. (1998)         \\
0,888.4365  &  $\pm$0.0002  &   2047.0  &   0.00037  &  I   &  Wolf et al. (1998)         \\
0,899.4721  &  $\pm$0.0003  &   2079.0  &   0.00000  &  I   &  Wolf et al. (1998)         \\
0,923.4410  &  $\pm$0.0017  &   2148.5  &   0.00015  &  II  &  Diethelm (1998)            \\
0,925.5099  &  $\pm$0.0014  &   2154.5  &  -0.00019  &  II  &  Diethelm (1998)            \\
0,926.3730  &  $\pm$0.0006  &   2157.0  &   0.00072  &  I   &  Bl\"attler (1998)          \\
0,927.4073  &  $\pm$0.0003  &   2160.0  &   0.00040  &  I   &  Wolf et al. (1998)         \\
1,270.5551  &  $\pm$0.0021  &   3155.0  &  -0.00137  &  I   &  Saf\'ar \& Zejda (2002)    \\
1,277.4539  &  $\pm$0.0025  &   3175.0  &  -0.00005  &  I   &  Saf\'ar \& Zejda (2002)    \\
1,284.3511  &  $\pm$0.0015  &   3195.0  &  -0.00033  &  I   &  Saf\'ar \& Zejda (2002)    \\
1,320.7360  &  $\pm$0.0008  &   3300.5  &   0.00038  &  II  &  Diethelm (2001)            \\
1,580.5980  &  $\pm$0.0033  &   4054.0  &  -0.00016  &  I   &  Zejda (2002)               \\
1,626.4677  &  $\pm$0.0015  &   4187.0  &   0.00121  &  I   &  Zejda (2002)               \\
1,641.2961  &  $\pm$0.0016  &   4230.0  &  -0.00002  &  I   &  Br\'at et al. (2007)       \\
1,684.4062  &  $\pm$0.0025  &   4355.0  &   0.00063  &  I   &  Br\'at et al. (2007)       \\
1,956.5138  &  $\pm$0.0015  &   5144.0  &   0.00039  &  I   &  Br\'at et al. (2007)       \\
1,965.4811  &  $\pm$0.0022  &   5170.0  &   0.00088  &  I   &  Zejda (2004)               \\
1,968.5846  &  $\pm$0.0025  &   5179.0  &   0.00049  &  I   &  Br\'at et al. (2007)       \\
2,014.4530  &  $\pm$0.0022  &   5312.0  &   0.00025  &  I   &  Br\'at et al. (2007)       \\
2,043.4225  &  $\pm$0.0022  &   5396.0  &   0.00009  &  I   &  Br\'at et al. (2007)       \\
2,053.4230  &  $\pm$0.0015  &   5425.0  &  -0.00084  &  I   &  Br\'at et al. (2007)       \\
2,287.5953  &  $\pm$0.0004  &   6104.0  &   0.00015  &  I   &  Bl\"attler (2002)          \\
2,344.4994  &  $\pm$0.0004  &   6269.0  &  -0.00039  &  I   &  Bl\"attler (2002)          \\
2,365.5369  &  $\pm$0.0032  &   6330.0  &  -0.00037  &  I   &  Zejda (2004)               \\
2,723.3454  &  $\pm$0.0009  &   7367.5  &  -0.00120  &  II  &  Diethelm (2003)            \\
2,723.5173  &  $\pm$0.0007  &   7368.0  &  -0.00174  &  I   &  Diethelm (2003)            \\
2,730.4160  &  $\pm$0.0049  &   7388.0  &  -0.00057  &  I   &  Zejda (2004)               \\
2,808.3591  &  $\pm$0.0003  &   7614.0  &   0.00048  &  I   &  Baki\c s et al. (2005)     \\
3,090.4682  &  $\pm$0.0001  &   8432.0  &   0.00077  &  I   &  Krajci (2005)              \\
3,094.4343  &  $\pm$0.0013  &   8443.5  &   0.00079  &  II  &  Diethelm (2004)            \\
3,117.5406  &  $\pm$0.0012  &   8510.5  &   0.00038  &  II  &  H\"ubscher et al. (2005)   \\
3,131.8527  &  $\pm$0.0003  &   8552.0  &   0.00012  &  I   &  Samec et al. (2006)        \\
3,135.8183  &  $\pm$0.0013  &   8563.5  &  -0.00036  &  II  &  Samec et al. (2006)        \\
3,143.2339  &  $\pm$0.0003  &   8585.0  &   0.00040  &  I   &  Krajci (2005)              \\
3,165.8238  &  $\pm$0.0003  &   8650.5  &   0.00091  &  II  &  Samec et al. (2006)        \\
3,203.4157  &  $\pm$0.0004  &   8759.5  &   0.00130  &  II  &  H\"ubscher et al. (2005)   \\
3,351.7109  &  $\pm$0.0002  &   9189.5  &  -0.00026  &  II  &  H\"ubscher et al. (2006)   \\
3,409.6501  &  $\pm$0.0044  &   9357.5  &  -0.00026  &  II  &  H\"ubscher et al. (2005)   \\
3,463.4508  &  $\pm$0.0002  &   9513.5  &  -0.00023  &  II  &  H\"ubscher et al. (2006)   \\
3,515.5263  &  $\pm$0.0010  &   9664.5  &  -0.00103  &  II  &  Diethelm (2005)            \\
3,767.9772  &  $\pm$0.0005  &  10396.5  &   0.00059  &  II  &  This article               \\
3,769.0115  &  $\pm$0.0003  &  10399.5  &   0.00026  &  II  &  This article               \\
3,770.0456  &  $\pm$0.0004  &  10402.5  &  -0.00027  &  II  &  This article               \\
3,797.9806  &  $\pm$0.0002  &  10483.5  &  -0.00022  &  II  &  This article               \\
3,813.4999  &  $\pm$0.0018  &  10528.5  &  -0.00034  &  II  &  H\"ubscher et al. (2006)   \\
3,821.7770  &  $\pm$0.0003  &  10552.5  &  -0.00027  &  II  &  Nelson (2007)              \\
3,859.5407  &  $\pm$0.0003  &  10662.0  &  -0.00049  &  I   &  Diethelm (2006)            \\
4,171.8267  &  $\pm$0.0001  &  11567.5  &   0.00038  &  II  &  This article               \\
4,172.8614  &  $\pm$0.0001  &  11570.5  &   0.00045  &  II  &  This article               \\
4,173.8959  &  $\pm$0.0001  &  11573.5  &   0.00032  &  II  &  This article               \\
4,174.7577  &  $\pm$0.0001  &  11576.0  &  -0.00007  &  I   &  This article               \\
4,176.8269  &  $\pm$0.0001  &  11582.0  &  -0.00012  &  I   &  This article               \\
4,177.8615  &  $\pm$0.0001  &  11585.0  &  -0.00015  &  I   &  This article               \\
4,521.0145  &  $\pm$0.0003  &  12580.0  &   0.00035  &  I   &  This article               \\
4,522.9109  &  $\pm$0.0002  &  12585.5  &  -0.00008  &  II  &  This article               \\
4,539.8093  &  $\pm$0.0002  &  12634.5  &  -0.00074  &  II  &  This article               \\
4,540.8442  &  $\pm$0.0002  &  12637.5  &  -0.00048  &  II  &  Nelson (2009)              \\
4,551.8804  &  $\pm$0.0003  &  12669.5  &  -0.00041  &  II  &  This article               \\
4,553.9497  &  $\pm$0.0003  &  12675.5  &  -0.00038  &  II  &  This article               \\
4,554.8127  &  $\pm$0.0002  &  12678.0  &   0.00042  &  I   &  This article               \\
4,554.9844  &  $\pm$0.0002  &  12678.5  &  -0.00032  &  II  &  This article               \\
4,555.5023  &  $\pm$0.0014  &  12680.0  &   0.00026  &  I   &  H\"ubscher et al. (2009)   \\
4,555.8471  &  $\pm$0.0002  &  12681.0  &   0.00019  &  I   &  This article               \\
4,556.8819  &  $\pm$0.0004  &  12684.0  &   0.00035  &  I   &  This article               \\
4,568.7795  &  $\pm$0.0002  &  12718.5  &  -0.00038  &  II  &  This article               \\
4,569.8142  &  $\pm$0.0002  &  12721.5  &  -0.00031  &  II  &  This article               \\
4,569.9873  &  $\pm$0.0003  &  12722.0  &   0.00035  &  I   &  This article               \\
4,598.4399  &  $\pm$0.0006  &  12804.5  &   0.00042  &  II  &  H\"ubscher et al. (2009)   \\
\enddata
\end{deluxetable}

\begin{deluxetable}{ccc}
\tablewidth{0pt}
\tablecaption{Parameters for the quadratic {\it plus} LTT ephemeris of AR Boo.$\rm ^a$}
\tablehead{
\colhead{Parameter}            &  \colhead{Values}            &  \colhead{Unit}
}
\startdata
$T_0$                          &  2,450,182.47724(28)         &  HJD                   \\
$P$                            &  0.344874257(32)             &  d                     \\
$A$                            &  1.264(27)$\times 10^{-11}$  &  d                     \\
$a_{12}\sin i_{3}$             &  0.259(67)                   &  AU                    \\
$\omega$                       &  280(18)                     &  deg                   \\
$e$                            &  0.59(28)                    &                        \\
$n  $                          &  0.1301(17)                  &  deg d$^{-1}$          \\
$T$                            &  2,446,142(97)               &  HJD                   \\
$P_{3}$                        &  7.573(99)                   &  yr                    \\
$K$                            &  0.00149(38)                 &  d                     \\
$f(M_{3})$                     &  0.000304(79)                &  $M_\odot$             \\
$M_3$ ($i_{3}$=90 deg)$\rm^b$  &  0.081                       &  $M_\odot$             \\
$M_3$ ($i_{3}$=60 deg)$\rm^b$  &  0.095                       &  $M_\odot$             \\
$M_3$ ($i_{3}$=30 deg)$\rm^b$  &  0.170                       &  $M_\odot$             \\
$dP$/$dt$                      &  2.676(58)$\times 10^{-7}$   &  d yr$^{-1}$           \\
$dM_2$/$dt$                    &  1.482$\times 10^{-7}$       &  $M_\odot$ yr$^{-1}$   \\
\enddata
\tablenotetext{a}{A parenthesized number is the 1$\sigma$-value of the last two digits of the associated parameter.}
\tablenotetext{b}{Masses of the hypothetical third body for different values of $i_{3}$.}
\end{deluxetable}

\begin{deluxetable}{cccc}
\tablewidth{0pt}
\tablecaption{Model parameters for possible magnetic activity of AR Boo.}
\tablehead{
\colhead{Parameter}       & \colhead{Primary}      & \colhead{Secondary}     & \colhead{Unit}
}
\startdata
$\Delta P$                & 0.1069                 &  0.1069                 &  s                   \\
$\Delta P/P$              & $3.59\times10^{-6}$    &  $3.59\times10^{-6}$    &                      \\
$\Delta Q$                & ${6.69\times10^{48}}$  &  ${1.72\times10^{49}}$  &  g cm$^2$            \\
$\Delta J$                & ${4.78\times10^{46}}$  &  ${8.67\times10^{46}}$  &  g cm$^{2}$ s$^{-1}$ \\
$I_{\rm s}$               & ${9.50\times10^{52}}$  &  ${5.78\times10^{53}}$  &  g cm$^{2}$          \\
$\Delta \Omega$           & ${5.03\times10^{-7}}$  &  ${1.50\times10^{-7}}$  &  s$^{-1}$            \\
$\Delta \Omega / \Omega$  & ${2.38\times10^{-3}}$  &  ${7.11\times10^{-4}}$  &                      \\
$\Delta E$                & ${4.80\times10^{40}}$  &  ${2.60\times10^{40}}$  &  erg                 \\
$\Delta L_{\rm rms}$      & ${6.31\times10^{32}}$  &  ${3.42\times10^{32}}$  &  erg s$^{-1}$        \\
                          & 0.162                  &  0.088                  &  $L_\odot$           \\
                          & 0.506                  &  0.144                  &  $L_{1,2}$           \\
$\Delta m_{\rm rms}$      & $\pm$0.174             &  $\pm$0.085             &  mag                 \\
$B$                       & 20640                  &  14576                  &  G
\enddata
\end{deluxetable}

\begin{deluxetable}{ccc}
\tablewidth{0pt}
\tablecaption{AR Boo parameters obtained by analyzing all LOAO light curves simultaneously.}
\tablehead{
\colhead{Parameter}             & \colhead{Primary}        & \colhead{Secondary}
}
\startdata
$T_0$ (HJD)                     & \multicolumn{2}{c}{2,453,767.8033(1)}                    \\
$P$ (d)                         & \multicolumn{2}{c}{0.34487694(6)}                        \\
$q$                             & \multicolumn{2}{c}{2.615(1)}                             \\
$i$ (deg)                       & \multicolumn{2}{c}{76.9(2)}                              \\
$T$ (K)                         & 5398(14)                 & 5100                          \\
$\Omega$                        & 6.036(6)                 & 6.102                         \\
$A$                             & 0.5                      & 0.5                           \\
$g$                             & 0.32                     & 0.32                          \\
$X$, $Y$                        & 0.645, 0.186             & 0.643, 0.169                  \\
$x_{B_{07}}$, $y_{B_{07}}$      & 0.945(51), 0.070         & 0.866(39), 0.006              \\
$x_{V_{07}}$, $y_{V_{07}}$      & 0.770(53), 0.183         & 0.854(44), 0.131              \\
$x_{R_{07}}$, $y_{R_{07}}$      & 0.605(55), 0.219         & 0.798(48), 0.186              \\
$x_{B_{08}}$, $y_{B_{08}}$      & 0.924(49), 0.070         & 0.808(40), 0.006              \\
$x_{V_{08}}$, $y_{V_{08}}$      & 0.737(51), 0.183         & 0.771(44), 0.131              \\
$x_{R_{08}}$, $y_{R_{08}}$      & 0.607(52), 0.219         & 0.699(48), 0.186              \\
$L/(L_1+L_2)_{B_{07}}$          & 0.376(2)                 & 0.624                         \\
$L/(L_1+L_2)_{V_{07}}$          & 0.371(2)                 & 0.629                         \\
$L/(L_1+L_2)_{R_{07}}$          & 0.366(2)                 & 0.634                         \\
$L/(L_1+L_2)_{B_{08}}$          & 0.372(2)                 & 0.628                         \\
$L/(L_1+L_2)_{V_{08}}$          & 0.366(2)                 & 0.634                         \\
$L/(L_1+L_2)_{R_{08}}$          & 0.356(2)                 & 0.644                         \\
$r$ (pole)                      & 0.2841(4)                & 0.4412(4)                     \\
$r$ (side)                      & 0.2968(5)                & 0.4725(6)                     \\
$r$ (back)                      & 0.3331(9)                & 0.5007(7)                     \\
$r$ (volume)$\rm ^a$            & 0.3064                   & 0.4729                        \\
$\Sigma W(O-C)^2$               & \multicolumn{2}{c}{0.0110}                               \\
\enddata
\tablenotetext{a}{Mean volume radius.}
\end{deluxetable}

\begin{deluxetable}{cccccccc}
\tabletypesize{\scriptsize}
\tablewidth{0pt}
\tablecaption{AR Boo parameters permitting $l_3$ and spots.}
\tablehead{
\colhead{Parameter}                     & \multicolumn{2}{c}{One-Spot Model} && \multicolumn{4}{c}{Two-Spot Model}  \\ [1.0mm] \cline{2-3} \cline{5-8} \\[-2.0ex]
                                        & \colhead{2007}    & \colhead{2008} && \multicolumn{2}{c}{2007}         & \multicolumn{2}{c}{2008}
}                                                                                                               
\startdata 
$\phi$$\rm ^a$                          & -0.00027(10)      & -0.00052(9)    && \multicolumn{2}{c}{-0.00144(9)}  & \multicolumn{2}{c}{-0.00038(9)} \\
$q$                                     & 2.601(7)          & 2.591(4)       && \multicolumn{2}{c}{2.580(4)}     & \multicolumn{2}{c}{2.590(5)}    \\
$i$ (deg)                               & 77.4(4)           & 77.6(4)        && \multicolumn{2}{c}{77.2(3)}      & \multicolumn{2}{c}{77.7(4)}     \\
$T_2$ (K)                               & 5388(16)          & 5378(14)       && \multicolumn{2}{c}{5382(15)}     & \multicolumn{2}{c}{5378(14)}    \\
$\Omega_1$=$\Omega_2$                   & 6.007(11)         & 5.982(9)       && \multicolumn{2}{c}{5.977(9)}     & \multicolumn{2}{c}{5.979(9)}    \\
$f$ (\%)                                & 12.3              & 14.3           && \multicolumn{2}{c}{12.5}         & \multicolumn{2}{c}{14.5}        \\
$x_{B1}$                                & 0.942(69)         & 0.966(57)      && \multicolumn{2}{c}{0.945(58)}    & \multicolumn{2}{c}{0.954(58)}   \\
$x_{V1}$                                & 0.756(68)         & 0.766(56)      && \multicolumn{2}{c}{0.767(58)}    & \multicolumn{2}{c}{0.748(57)}   \\
$x_{R1}$                                & 0.575(68)         & 0.627(58)      && \multicolumn{2}{c}{0.601(58)}    & \multicolumn{2}{c}{0.606(58)}   \\
$x_{B2}$                                & 0.857(92)         & 0.849(67)      && \multicolumn{2}{c}{0.823(73)}    & \multicolumn{2}{c}{0.853(63)}   \\
$x_{V2}$                                & 0.793(85)         & 0.789(68)      && \multicolumn{2}{c}{0.801(73)}    & \multicolumn{2}{c}{0.785(64)}   \\
$x_{R2}$                                & 0.700(84)         & 0.704(71)      && \multicolumn{2}{c}{0.751(74)}    & \multicolumn{2}{c}{0.698(66)}   \\
$L_1$/($L_{1}$+$L_{2}$+$L_{3}$){$_{B}$} & 0.361(4)          & 0.360(3)       && \multicolumn{2}{c}{0.363(3)}     & \multicolumn{2}{c}{0.360(3)}    \\
$L_1$/($L_{1}$+$L_{2}$+$L_{3}$){$_{V}$} & 0.359(4)          & 0.359(3)       && \multicolumn{2}{c}{0.362(3)}     & \multicolumn{2}{c}{0.359(3)}    \\
$L_1$/($L_{1}$+$L_{2}$+$L_{3}$){$_{R}$} & 0.357(4)          & 0.352(4)       && \multicolumn{2}{c}{0.359(3)}     & \multicolumn{2}{c}{0.352(4)}    \\
{\it $l_{3B}$}$\rm ^{b}$                & 0.036(18)         & 0.025(14)      && \multicolumn{2}{c}{0.020(14)}    & \multicolumn{2}{c}{0.028(13)}   \\
{\it $l_{3V}$}$\rm ^{b}$                & 0.018(17)         & 0.012(14)      && \multicolumn{2}{c}{0.012(14)}    & \multicolumn{2}{c}{0.015(14)}   \\
{\it $l_{3R}$}$\rm ^{b}$                & 0.006(17)         & 0.007(14)      && \multicolumn{2}{c}{0.010(14)}    & \multicolumn{2}{c}{0.009(14)}   \\
$r_1$ (pole)                            & 0.2853(10)        & 0.2865(8)      && \multicolumn{2}{c}{0.2860(8)}    & \multicolumn{2}{c}{0.2866(8)}   \\
$r_1$ (side)                            & 0.2981(12)        & 0.2995(9)      && \multicolumn{2}{c}{0.2989(9)}    & \multicolumn{2}{c}{0.2997(10)}  \\
$r_1$ (back)                            & 0.3350(21)        & 0.3371(16)     && \multicolumn{2}{c}{0.3358(16)}   & \multicolumn{2}{c}{0.3373(17)}  \\
$r_1$ (volume)                          & 0.3079            & 0.3095         && \multicolumn{2}{c}{0.3087}       & \multicolumn{2}{c}{0.3097}      \\
$r_2$ (pole)                            & 0.4414(8)         & 0.4420(7)      && \multicolumn{2}{c}{0.4408(7)}    & \multicolumn{2}{c}{0.4420(7)}   \\
$r_2$ (side)                            & 0.4728(11)        & 0.4736(9)      && \multicolumn{2}{c}{0.4721(9)}    & \multicolumn{2}{c}{0.4737(9)}   \\
$r_2$ (back)                            & 0.5014(15)        & 0.5026(12)     && \multicolumn{2}{c}{0.5008(12)}   & \multicolumn{2}{c}{0.5027(12)}  \\
$r_2$ (volume)                          & 0.4734            & 0.4743         && \multicolumn{2}{c}{0.4728}       & \multicolumn{2}{c}{0.4744}      \\
\multicolumn{8}{l}{Spot parameters:}                                                                                                               \\
Colatitude$_2$ (deg)                    & 93.4(2.6)         & 90.4(0.6)      && 91.5(2.8)       & 103.0(2.4)     & 83.2(1.0)       & 103.5(1.5)    \\
Longitude$_2$ (deg)                     & 172.5(1.3)        & 97.4(1.3)      && 160.4(0.4)      & 39.7(2.8)      & 89.4(0.4)       & 64.6(1.1)     \\
Radius$_2$ (deg)                        & 14.99(46)         & 16.07(21)      && 14.42(35)       & 17.25(92)      & 17.71(18)       & 14.63(73)     \\
$T$$\rm _{spot_2}$/$T$$\rm _{local_2}$  & 0.928(3)          & 0.917(3)       && 0.925(2)        & 1.033(2)       & 0.918(2)        & 1.024(3)      \\
$\Sigma W(O-C)^2$                       & 0.0068            & 0.0084         && \multicolumn{2}{c}{0.0058}       & \multicolumn{2}{c}{0.0084}      
\enddata
\tablenotetext{a}{Phase shift from the data phased by the linear ephemeris of Table 5.}
\tablenotetext{b}{Value at 0.75 phase.}
\end{deluxetable}

\begin{deluxetable}{cccccccccc}
\tabletypesize{\scriptsize}
\tablewidth{0pt}
\tablecaption{AR Boo parameters obtained from historical light curves.}
\tablehead{
\colhead{Parameter}                      & \multicolumn{4}{c}{Samec et al. (2006)}                                  && \multicolumn{4}{c}{Wolf et al. (1998)}                                  \\ [1.0mm] \cline{2-5} \cline{7-10} \\[-1.0ex]
                                         & \multicolumn{2}{c}{Primary}           & \multicolumn{2}{c}{Secondary}    && \multicolumn{2}{c}{Primary}           & \multicolumn{2}{c}{Secondary}   
}                                                                                                                   
\startdata                                                                                                          
$T_0$ (HJD)                              & \multicolumn{4}{c}{2,450,182.480(15)}                                    && \multicolumn{4}{c}{2,450,182.4793(27)}                                  \\
$P$ (day)                                & \multicolumn{4}{c}{0.3448752(18)}                                        && \multicolumn{4}{c}{0.3448733(21)}                                       \\
$q$                                      & \multicolumn{4}{c}{2.582(15)}                                            && \multicolumn{4}{c}{2.580(13)}                                           \\
$i$ (deg)                                & \multicolumn{4}{c}{77.1(1.0)}                                            && \multicolumn{4}{c}{77.69(93)}                                           \\
$T$ (K)                                  & \multicolumn{2}{c}{5384(54)}          & \multicolumn{2}{c}{5100}         && \multicolumn{2}{c}{5383(55)}          & \multicolumn{2}{c}{5100}        \\
$\Omega$                                 & \multicolumn{2}{c}{5.975(31)}         & \multicolumn{2}{c}{5.975}        && \multicolumn{2}{c}{5.977(25)}         & \multicolumn{2}{c}{5.975}       \\
$f$ (\%)                                 & \multicolumn{4}{c}{13.4}                                                 && \multicolumn{4}{c}{12.5}                                                \\                
$x_{B}$                                  & \multicolumn{2}{c}{0.94(21)}          & \multicolumn{2}{c}{0.88(23)}     && \multicolumn{2}{c}{\dots}             & \multicolumn{2}{c}{\dots}       \\
$x_{V}$                                  & \multicolumn{2}{c}{0.63(21)}          & \multicolumn{2}{c}{0.83(25)}     && \multicolumn{2}{c}{0.72(16)}          & \multicolumn{2}{c}{0.88(15)}    \\
$x_{R}$                                  & \multicolumn{2}{c}{0.51(23)}          & \multicolumn{2}{c}{0.74(27)}     && \multicolumn{2}{c}{0.60(13)}          & \multicolumn{2}{c}{0.75(13)}    \\
$x_{I}$                                  & \multicolumn{2}{c}{0.38(24)}          & \multicolumn{2}{c}{0.64(29)}     && \multicolumn{2}{c}{\dots}             & \multicolumn{2}{c}{\dots}       \\
$L_1$/($L_{1}$+$L_{2}$+$L_{3}$){$_{B}$}  & \multicolumn{2}{c}{0.370(12)}         & \multicolumn{2}{c}{0.611}        && \multicolumn{2}{c}{\dots}             & \multicolumn{2}{c}{\dots}       \\
$L_1$/($L_{1}$+$L_{2}$+$L_{3}$){$_{V}$}  & \multicolumn{2}{c}{0.376(12)}         & \multicolumn{2}{c}{0.605}        && \multicolumn{2}{c}{0.374(9)}          & \multicolumn{2}{c}{0.613}       \\
$L_1$/($L_{1}$+$L_{2}$+$L_{3}$){$_{R}$}  & \multicolumn{2}{c}{0.366(13)}         & \multicolumn{2}{c}{0.623}        && \multicolumn{2}{c}{0.361(8)}          & \multicolumn{2}{c}{0.634}       \\
$L_1$/($L_{1}$+$L_{2}$+$L_{3}$){$_{I}$}  & \multicolumn{2}{c}{0.355(13)}         & \multicolumn{2}{c}{0.626}        && \multicolumn{2}{c}{\dots}             & \multicolumn{2}{c}{\dots}       \\
{\it $l_{3B}$}$\rm ^{a}$                 & \multicolumn{4}{c}{0.019(50)}                                            && \multicolumn{4}{c}{\dots}                                               \\
{\it $l_{3V}$}$\rm ^{a}$                 & \multicolumn{4}{c}{0.019(50)}                                            && \multicolumn{4}{c}{0.013(35)}                                           \\
{\it $l_{3R}$}$\rm ^{a}$                 & \multicolumn{4}{c}{0.011(51)}                                            && \multicolumn{4}{c}{0.005(32)}                                           \\
{\it $l_{3I}$}$\rm ^{a}$                 & \multicolumn{4}{c}{0.019(50)}                                            && \multicolumn{4}{c}{\dots}                                               \\
$r$ (pole)                               & \multicolumn{2}{c}{0.2863(26)}        & \multicolumn{2}{c}{0.4413(23)}   && \multicolumn{2}{c}{0.2860(22)}        & \multicolumn{2}{c}{0.4408(19)}  \\
$r$ (side)                               & \multicolumn{2}{c}{0.2993(32)}        & \multicolumn{2}{c}{0.4728(30)}   && \multicolumn{2}{c}{0.2989(26)}        & \multicolumn{2}{c}{0.4721(25)}  \\
$r$ (back)                               & \multicolumn{2}{c}{0.3366(54)}        & \multicolumn{2}{c}{0.5016(40)}   && \multicolumn{2}{c}{0.3358(45)}        & \multicolumn{2}{c}{0.5008(33)}  \\
$r$ (volume)                             & \multicolumn{2}{c}{0.3092}            & \multicolumn{2}{c}{0.4735}       && \multicolumn{2}{c}{0.3087}            & \multicolumn{2}{c}{0.4728}      \\
Colatitude (deg)$\rm ^{b}$               & \multicolumn{2}{c}{\dots}             & 90.7(13.0)        & 115.3(2.9)   && \multicolumn{2}{c}{\dots}             & 79.9(0.8)         & 98.9(9.4)   \\
Longitude (deg)$\rm ^{b}$                & \multicolumn{2}{c}{\dots}             & 150.3(2.9)        & 35.7(2.5)    && \multicolumn{2}{c}{\dots}             & 106.7(0.7)        & 298(11.0)   \\
Radius (deg)$\rm ^{b}$                   & \multicolumn{2}{c}{\dots}             & 11.4(1.4)         & 19.0(0.8)    && \multicolumn{2}{c}{\dots}             & 20.8(0.4)         & 13.1(2.9)   \\
$T$$\rm _{spot}$/$T$$\rm _{local}$$\rm ^{b}$  & \multicolumn{2}{c}{\dots}        & 0.931(0.012)      & 1.086(0.006) && \multicolumn{2}{c}{\dots}             & 0.870(0.006)      & 1.024(0.010)   
\enddata
\tablenotetext{a}{Value at 0.75 phase.}
\tablenotetext{b}{The errors for these parameters are given in full.}
\end{deluxetable}

\begin{deluxetable}{ccc}
\small
\tablewidth{195pt}
\tablecaption{Estimated absolute dimensions of AR Boo.}
\tablehead{
\colhead{Parameter} & \colhead{Primary} & \colhead{Secondary}}
\startdata
$M$ ($M_\odot$)     &  0.35         &  0.90           \\
$R$ ($R_\odot$)     &  0.65         &  1.00           \\
$\log$ $g$ (cgs)    &  4.36         &  4.39           \\
$L$ ($L_\odot$)     &  0.32         &  0.61           \\
$M_{\rm bol}$ (mag) &  5.94         &  5.23           \\
BC (mag)            &  $-$0.17      &  $-$0.26        \\
$M_{V}$ (mag)       &  6.11         &  5.49           \\
Distance (pc)       &  \multicolumn{2}{c}{352}        \\
\enddata
\end{deluxetable}

\begin{deluxetable}{ccccccl}
\tablewidth{0pt}
\tablecaption{Minimum timings determined by the WD code from individual eclipses of AR Boo.}
\tablehead{
\colhead{Observed$\rm^{a,b}$} & \colhead{WD$\rm^{b}$} & \colhead{Error$\rm^{c}$} & \colhead{Difference$\rm^{d}$} & \colhead{Filter} & \colhead{Min} & References
}
\startdata
0,607.3635  &  0,607.36271  &  $\pm$0.00021  &  $+$0.00079  &  $VR$    &  I   &  Wolf et al.    \\
0,607.5346  &  0,607.53490  &  $\pm$0.00022  &  $-$0.00030  &  $VR$    &  II  &  Wolf et al.    \\
0,611.5025  &  0,611.50142  &  $\pm$0.00028  &  $+$0.00108  &  $R$     &  I   &  Wolf et al.    \\
0,638.4028  &  0,638.40188  &  $\pm$0.00013  &  $+$0.00092  &  $R$     &  I   &  Wolf et al.    \\
3,131.8527  &  3,131.85333  &  $\pm$0.00016  &  $-$0.00063  &  $BVRI$  &  I   &  Samec et al.   \\
3,135.8183  &  3,135.81836  &  $\pm$0.00016  &  $-$0.00060  &  $BVRI$  &  II  &  Samec et al.   \\
3,165.8238  &  3,165.82275  &  $\pm$0.00020  &  $+$0.00105  &  $BVRI$  &  II  &  Samec et al.   \\
4,171.8267  &  4,171.82622  &  $\pm$0.00006  &  $+$0.00048  &  $BVR$   &  II  &  This article   \\
4,172.8614  &  4,172.86085  &  $\pm$0.00009  &  $+$0.00055  &  $BVR$   &  II  &  This article   \\
4,173.8959  &  4,173.89546  &  $\pm$0.00005  &  $+$0.00044  &  $BVR$   &  II  &  This article   \\
4,174.7577  &  4,174.75757  &  $\pm$0.00007  &  $+$0.00013  &  $BVR$   &  I   &  This article   \\
4,176.8269  &  4,176.82668  &  $\pm$0.00008  &  $+$0.00022  &  $BVR$   &  I   &  This article   \\
4,177.8615  &  4,177.86136  &  $\pm$0.00007  &  $+$0.00014  &  $BVR$   &  I   &  This article   \\
4,521.0145  &  4,521.01425  &  $\pm$0.00011  &  $+$0.00025  &  $BVR$   &  I   &  This article   \\
4,522.9109  &  4,522.91122  &  $\pm$0.00012  &  $-$0.00032  &  $BVR$   &  II  &  This article   \\
4,539.8093  &  4,539.80960  &  $\pm$0.00012  &  $-$0.00030  &  $BVR$   &  II  &  This article   \\
4,551.8804  &  4,551.88093  &  $\pm$0.00009  &  $-$0.00053  &  $BVR$   &  II  &  This article   \\
4,553.9497  &  4,553.95013  &  $\pm$0.00015  &  $-$0.00043  &  $BVR$   &  II  &  This article   \\
4,554.8127  &  4,554.81233  &  $\pm$0.00009  &  $+$0.00037  &  $BVR$   &  I   &  This article   \\
4,554.9844  &  4,554.98474  &  $\pm$0.00012  &  $-$0.00034  &  $BVR$   &  II  &  This article   \\
4,555.8471  &  4,555.84711  &  $\pm$0.00011  &  $-$0.00001  &  $BVR$   &  I   &  This article   \\
4,556.8819  &  4,556.88168  &  $\pm$0.00015  &  $+$0.00022  &  $BVR$   &  I   &  This article   \\
4,568.7795  &  4,568.77993  &  $\pm$0.00009  &  $-$0.00043  &  $BVR$   &  II  &  This article   \\
4,569.8142  &  4,569.81443  &  $\pm$0.00016  &  $-$0.00023  &  $BVR$   &  II  &  This article   \\
4,569.9873  &  4,569.98727  &  $\pm$0.00011  &  $+$0.00003  &  $BVR$   &  I   &  This article   \\
\enddata
\tablenotetext{a}{cf. Table 2.}
\tablenotetext{b}{HJD 2,450,000 is suppressed.}
\tablenotetext{c}{Uncertainties yielded by the WD code.}
\tablenotetext{d}{Differences between columns (1) and (2).}
\end{deluxetable}

\end{document}